\documentclass{elsart}
\usepackage{graphicx,harvard,amssymb}

\makeatletter
\def\elsartstyle{%
	\def\normalsize{\@setfontsize\normalsize\@xiipt{14.5}}
	\def\small{\@setfontsize\small\@xipt{13.6}}
	\let\footnotesize=\small
	\def\large{\@setfontsize\large\@xivpt{18}}
	\def\Large{\@setfontsize\Large\@xviipt{22}}
	\skip\@mpfootins = 18\p@ \@plus 2\p@
	\normalsize
}
\makeatother
\def\Msun{{\rm\,M_\odot}}
\def\simlt{\mathrel{\spose{\lower 3pt\hbox{$\mathchar"218$}}
     \raise 2.0pt\hbox{$\mathchar"13C$}}}
\def\simgt{\mathrel{\spose{\lower 3pt\hbox{$\mathchar"218$}}
     \raise 2.0pt\hbox{$\mathchar"13E$}}}
\def\spose#1{\hbox to 0pt{#1\hss}}
\def\lta{\mathrel{\spose{\lower 3pt\hbox{$\mathchar"218$}}
     \raise 2.0pt\hbox{$\mathchar"13C$}}}
\def\gta{\mathrel{\spose{\lower 3pt\hbox{$\mathchar"218$}}
     \raise 2.0pt\hbox{$\mathchar"13E$}}}
\def\approxlt{\mathrel{\spose{\lower 3pt\hbox{$\sim$}}
	\raise 2.0pt\hbox{$<$}}}
\def\approxgt{\mathrel{\spose{\lower 3pt\hbox{$\sim$}}
	\raise 2.0pt\hbox{$>$}}}
\def\approxpropto{\mathrel{\spose{\lower 3pt\hbox{$\sim$}}
	\raise 2.0pt\hbox{$\propto$}}}

\def\bibcode#1{(\texttt{#1})}

\def\url#1{{\ttfamily\def\/{/\discretionary{}{}{}}#1}}

\def\et{{\it et~al. }}

\begin{document}

\begin{frontmatter}
\title{How to find MACHOs in the Virgo Cluster}
\author[suss]{Helen Tadros\thanksref{htemail}}
\author[imp]{Stephen Warren\thanksref{sjwemail}}
\author[cam]{Paul Hewett\thanksref{phemail}}
\thanks[htemail]{E-mail: helent@astr.cpes.susx.ac.uk}
\thanks[sjwemail]{E-mail: s.j.warren@ic.ac.uk}
\thanks[phemail]{E-mail: phewett@ast.cam.ac.uk}
\address[suss]{Astronomy Centre, Sussex University, Falmer, Brighton, UK, BN1 9QH}
\address[imp]{Astrophysics, Blackett Labs, ICSTM, Prince Consort Rd, London SW7 2BZ}
\address[cam]{Institute of Astronomy, Madingley Rd. Cambridge, UK, CB3 OHA}

\begin{abstract}
We discuss the feasibility of finding extra-galactic MACHOs by
monitoring quasars behind the Virgo cluster of galaxies. We show that
with only a modest observing programme one could detect several MACHOs in
the mass range $1\times10^{-5}-2\times 10^{-2}\Msun$ if they make a
significant contribution to the mass of Virgo. The contamination by
events from cosmologically distributed MACHOs is estimated and is
negligible if either the MACHO mass is $\gta 10^{-4}\Msun$ or the quasar
radius is $\gta 3\times 10^{15}$ cm.

\end{abstract}

\end{frontmatter}

\section{Introduction} The optical depth to microlensing toward the LMC
is a few $\times 10^{-7}$ and the detection of halo microlensing events
requires extensive monitoring campaigns. The MACHO project, for
example, after monitoring several million stars over a $2.3$ year
period reported only $8$ events towards the LMC.  In this article we
discuss a search for MACHOs in the Virgo cluster, based on the
microlensing of background quasars.  As shown below the mean optical
depth over the central 10 deg$^2$ degrees of Virgo is $\sim
2\times10^{-3}$.  Therefore one need monitor only hundreds of
background quasars $-$ compared with millions of stars in the LMC.  For
a relatively small investment in telescope time it is possible to
obtain important new information on the nature of dark matter in rich
clusters.

\section{The Optical Depth to Microlensing in Virgo}
The case for microlensing searches in clusters was first made in an
excellent paper by \citeasnoun{WI}. 
In this section we calculate the
optical depth to microlensing of background quasars through the Virgo
cluster. The motivation for choosing Virgo is explained below. We
assume that Virgo can be modelled as an isothermal sphere of 1--D
velocity dispersion $\sigma_{_{1D}}=670$ km s$^{-1}$ \cite{danese}. The
effect of the core radius can be neglected. The quasars (e.g. at $z\sim
1$) are much more distant than Virgo (at 16 Mpc) and in this limit
$D_s\gg D_d$ the optical depth at angle $\theta$ (in degrees) from the
line of sight to the cluster centre is given by
$\tau=360\sigma_{_{1D}}^2/(\theta c^2)=1.8\times10^{-3}/\theta$.
Therefore the average optical depth over the central 10 square degrees
of the Virgo cluster is $\sim 2.0\times10^{-3}$.

\section{The effect of source size}
We now calculate the effect of finite source size.  \citeasnoun{WI}
appear to have plotted this incorrectly (their Figure 3) and
this has consequences for the optimal observing strategy. Here we
suppose that once the radius of the Einstein ring (projected in the
source plane) is less than half the quasar radius microlensing events
will be undetectable since the light curve will be severely distorted
relative to the point--mass approximation, and the maximum
amplification, for perfect alignment, is $<\sqrt2$. The typical size of
the continuum--emitting region in a quasar is still not well
known. Theoretical considerations \cite{Rees} and one observational
measurement \cite{wambs} suggest that quasar radii lie in the range
$10^{14}$ -- $10^{16}$cm, corresponding to $3-30$ Schwarzschild radii
for black--hole masses from $10^8-10^9\Msun$.

\begin{figure}
\centering
\begin{picture}(200,210)
\includegraphics{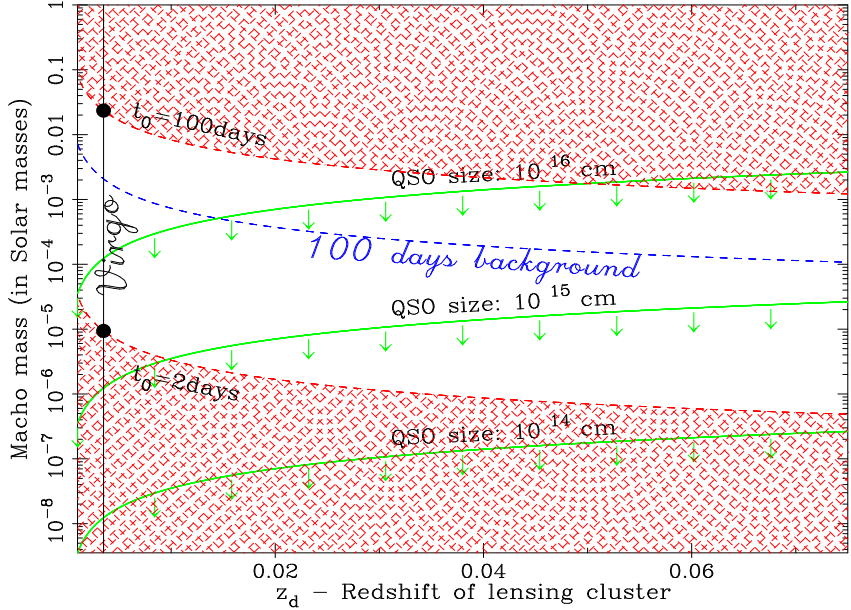}
\end{picture}
\caption{\label{fig2} Plot showing the range of MACHO masses to which
our example programme (as specified in the text) is sensitive. The
shaded regions are excluded by the length of the monitoring
programme. For a given quasar size regions below the solid lines are
excluded due to the effect of finite source size.  The curve marked
{\it 100 days background} is discussed in \S4}
\end{figure}

We consider now a 150--day programme of nightly monitoring of quasars
behind a massive cluster of velocity dispersion equal to that of the
Virgo cluster.  Figure~\ref{fig2} plots the range of MACHO masses
detectable in such a survey as a function of redshift of the
cluster. We assume a typical transverse MACHO velocity of $\sqrt2
\sigma_{_{1D}}$.  The shaded regions are inaccessible because we would
not see light curves with time-scales longer than $\sim 100$ days or
shorter than $\sim 2$ days.  The solid lines with the arrows show
the constraints introduced by the size of the quasar for three possible
quasar radii. For a given radius regions below the line are excluded.
It can be seen that as the redshift of the cluster increases the
accessible range of MACHO mass is squeezed: from above by the duration
of the survey, and from below by the quasar source size. At lower
redshifts the effect of finite source size is minimised, so we want to
target the nearest massive cluster in order to maximise the range of
masses detectable. This is the reason for choosing Virgo, as well as the
fact that higher masses are reached. If quasar radii are
$<3\times10^{15}$cm we could detect MACHOs in the mass range $1 \times
10^{-5} - 2 \times 10^{-2} \Msun$. The other suitable target is the
Perseus cluster ($z=0.018$, $\sigma_{_{1D}}=1010$). The optical depth
through Perseus is more than twice as large as through Virgo, at the
expense of a smaller range of masses explored.

We plan to monitor quasars brighter than $R=20.3$ at which point the
surface density of quasars is 30 deg$^{-2}$. Therefore we will obtain
light curves for about 300 quasars and the probability that a
microlensing event is taking place at any one time is $0.60\alpha$
where $\alpha$ is the fraction of the mass of the cluster in MACHOs.
If all the dark matter were in MACHOs of mass $1 \times 10^{-5} \Msun$
(corresponding to the shortest observable time-scale of $\sim 2$ days)
we would expect to see $ 0.60 \times 150/2 = 45$ events. If the typical
MACHO mass were $2 \times 10^{-2} \Msun$ (corresponding to the longest
observable time-scale of $\sim 100$ days) we would see $0.60 \times
150/100 \sim 1 $ event. Thus in $150$ days we could measure or place
interesting limits on the fraction of the mass of Virgo in MACHOs, over
a mass range of over 3 orders of magnitude.

\section{Contamination by cosmologically--distributed MACHOs}

Another important issue is the likely contamination by events due to
foreground and background MACHOs. If a substantial fraction of the mass
of Virgo is in MACHOs the same is presumably true on cosmological
scales. Nevertheless the velocity dispersion of the
cosmologically--distributed MACHOs will be much less than that for
Virgo. We will assume $\sigma_{_{1D}}=200$ km s$^{-1}$ for `field'
MACHOs. The detection limit corresponding to a 100--day event is
plotted as the dashed line in Figure~\ref{fig2}, so that field MACHOs
above this line will not be detected. The same is true of field MACHOs
lying below whichever is the correct quasar--limit line. We see that
field MACHOs can be found out to a limiting redshift $z_{lim}$, where
$z_{lim}$ is the redshift where a horizontal line (for that mass) first
cuts one of these detection limits.  The optical depth to microlensing
for cosmological compact objects is $ \sim \Omega_{_{M}} z_{lim}^{2}/4$
for $z_{lim} \ll 1$ \cite{PG} so that the optical depth can be computed
as a function of MACHO mass, for different quasar source sizes and
different values of the cosmological density in MACHOs $\Omega_{_{M}}$.
For example suppose that $\Omega_{_{M}}=0.2$. We find that if the MACHO
mass is $\gta 10^{-4}$ the optical depth of field
MACHOs is less than $10\%$ of the average optical depth through Virgo,
regardless of the quasar size. Field MACHOs are similarly
irrelevant if the quasar radii are $\gta 3\times 10^{15}$
cm, regardless of MACHO mass.  On the other hand for masses less than
$\sim10^{-4}$ if the linear sizes of quasars are much less than $1
\times 10^{15}$cm then we may have an appreciable cosmological
background of events.

\section{Further remarks}
Besides the large optical depth there are several other advantages to
using clusters as the target for microlensing searches. Firstly, the
mass of the lensing object will be better constrained than in existing
microlensing searches as the distance to the lens is known (although
see \S4). Secondly one avoids the problem of blending of the sources
which complicates the current microlensing searches. The intrinsic
variability of the background quasars is relevant, but should not be a
serious problem as most optical quasars show only modest variability on
the time-scales $-$ a few days to a few months $-$ in which we are
interested \cite{hook}. The $rms$ intrinsic magnitude variation of a
quasar over a period of $\sim 3$ months is $0.09$ mag.  \cite{Warren},
while we will be looking for magnifications of $> 0.3$ mag. In any case
it is the the one-off nature and the characteristic time profile of
microlensing events which are their signature and which will be the
primary means of distinguishing between microlensing and intrinsic
variability.

We have shown how it is feasible to obtain direct limits on, or make
detections of, extra-galactic compact baryonic dark matter. Currently
we are only able to probe the nature of this type of dark matter on
scales corresponding to the Galaxy halo (of order tens of kpc). A
project such as the one described here would obtain information on the
nature of dark matter on scales of order 1Mpc. We plan to modify the
observing strategy to make it suitable for use with the INT wide field
camera. To quantify our selection effects we also plan to make a
detailed study of the effects of intrinsic quasar variability on the
detection efficiency of MACHOs as a function of mass (Tadros, Hewett
and Warren {\it{in prep.}}).

{\it{\underline{Question from Dr. Mike Hawkins:} Given that there are
probably ways of distinguishing intrinsic variation from microlensing,
how would you separate microlensing in the Virgo cluster from more
general microlensing along the line of sight to the quasars. }} \\

{\it{Answer}}\\ This is an important issue and the question prompted us
to quantify the contamination due to cosmological MACHOs (see \S4). The
expected background event rate depends critically on the quasar
size. It should be possible to determine whether we are primarily
monitoring events in Virgo or more generally along the line of sight
from a combination of (a) the statistical distribution of events over
the field of view and by comparison with a control field, (b) an
examination of the timescales, (c) the possible measurement of the
quasar sizes in events which show flat--topped light curves, because
the source angular size is comparable to the Einstein angle.  Of course
one is just using Virgo as a convenient testing ground, and we would like
to know the nature of dark matter throughout the Universe. Detection of
MACHOs outside of the Virgo cluster is therefore of great importance.

\end{document}